\documentstyle[twocolumn,aps,prl,psfig]{revtex}

\begin{document}

\title{ Phase diffusion of Bose-Einstein condensates in a one-dimensional optical
lattice}

\author{Hongwei Xiong$^{1}$, Shujuan Liu$^{1}$, Guoxiang Huang$^{2}$, and Lei Wang$^{1}$}
\address{$^{1}$Department of Applied Physics, Zhejiang
University of Technology, Hangzhou 310032, China}
\address{$^{2}$Department of Physics and Key Laboratory for Optical and Magnetic
Resonance Spectroscopy, East China Normal University, Shanghai
200062, China}

\date{\today}
\maketitle

\begin{abstract}
We investigate the phase diffusion of a Bose-Einstein condensate
(BEC) confined in the combined potential of a magnetic
trap and a one-dimensional optical lattice. We show that the phase
diffusion of the condensate in the whole optical lattice is
evident and can be measured from the interference pattern of
expanding subcondensates after the optical lattice is
switched off. The maximum density of the interference pattern
decreases significantly due to the mixing of the phase diffusion
appearing in different subcondensates. This suggests a way to
detect experimentally the notable phase diffusion of BECs.
\end{abstract}


\narrowtext

%%%%%%%%%%%%%%%%%%%%%%%%%%%%%%%%%%%%%%%%%%%%%%%%%%%%%

%%%%%%%%%%%%%%%%%%%%%%%%%%%%%%%%%%%%%%%%%%%%%%%%%%%%%

\section{Introduction}

The remarkable experimental realization of Bose-Einstein condensation has
generated intensive theoretical and experimental investigations for weakly
interacting Bose gases\cite{RMP,NATURE}. Although much progress has been
made, one of the challenging problems to be solved in both theory and
experiment is how to detect the phase diffusion of a Bose-Einstein
condensate (BEC). The study of the phase diffusion is of crucial importance
because the number of particles realized in all BEC experiments is finite.
Within the formalism of the mean-field theory, it is known that the
condensate can be well described by a macroscopic wave function (or called
an order parameter) with a single phase \cite{RMP}. The development of the
phase at low temperature is governed by the Gross-Pitaevskii (GP) equation.
However, for a finite system and hence for trapped Bose gases, although a
macroscopic wave function describing a BEC still exists, as is now firmly
established experimentally, the phase of the macroscopic wave function will
undergo a quantum diffusion process as time develops\cite{LEW}. Because the
phase diffusion reflects directly the coherent nature of a condensate, the
study of the phase diffusion of BECs is not only fundamental but also
important for applications such as the realization of an atom laser\cite{NAR}%
.

For a condensate confined in a trap, the development of its phase
is determined by the chemical potential of the system. Due to
quantum and thermal fluctuations, the phase diffusion is described
by the fluctuations of the chemical potential. Even at initial
state the condensate has an exact single phase, the phase will
become unpredictable due to the fluctuations of the condensate.
After the realization of Bose-Einstein condensation, much
theoretical attention has been paid to the phase diffusion of
BECs\cite
{LEW,NAR,YOU,JAV,MOLMER,GRAHAM,LEGGETT,JAK,DALVIT,KUKLOV,XIONG}.
Because for a single condensate it is hard to observe the effect
of phase diffusion based on the density distribution of a
Bose-condensed gas, in the experiment conducted by the JILA
group\cite{HALL} two expanding condensates are used to produce an
interference pattern and hence the phase diffusion of the
condensates can be detected. It was found that in this experiment
there is no detectable phase diffusion on the time scale up to
$100$ ${\rm ms}$.

Recently, the BECs in optical lattices have attracted increasing attention
(see Refs \cite{AND,ORZ,GRE1,CAT,GREI} and references therein). Trapping
atoms in optical lattices can provide very effective and powerful means for
controlling the properties of a BEC, especially the quantum phase effects on
a macroscopic scale such as the coherence of matter waves \cite{GRE1,GREI}.
In the present work, we investigate the effect of the phase diffusion of a
BEC confined in the combined potential of a harmonically magnetic trap and a
one-dimensional (1D) optical lattice. In the presence of the 1D optical
lattice, there are an array of disk-shaped subcondensates. For each
subcondensate, i. e. the BEC formed in the corresponding potential well
induced by the optical lattice, there exist phase fluctuations which
increase very slowly with the development of time. For high enough
potential-well intensity, the phase fluctuations for different
subcondensates are independent from each other. But when the potential-well
intensity is not high, there will be a correlation for the subcondensates in
different wells and hence the situation may be quite different. One expects
that the phase diffusion of the BEC in the whole optical lattice can display
notable new characters because of the mixing effect of the phase diffusion
appearing in different subcondensates.

The paper is organized as follows. Section II is devoted to the phase
diffusion of a single 2D (as well as quasi-2D) condensate for the
temperature far below the critical temperature. In Sec. III, the phase
diffusion of an array of disk-shaped subcondensates induced by a 1D optical
lattice is investigated based on the interference pattern resulting from a
superposition of expanding subcondensates in different potential wells for
different holding time of the optical lattice. Sec. IV contains a discussion
and summary of our results.

\section{Phase diffusion of a quasi-2D condensate}

We consider an array of subcondensates formed in the combined potential
consisting of a harmonically magnetic trap and a periodic optical lattice.
The combined potential is described by \cite{1Dlatt,MORSCH,MORSCH1}:
\begin{equation}
V=\frac{1}{2}m\omega _{x}^{2}x^{2}+\frac{1}{2}m\omega _{\bot }^{2}\left(
y^{2}+z^{2}\right) +sE_{R}\sin ^{2}\left( \frac{2\pi x}{\lambda }\right) ,
\label{potential}
\end{equation}
where $\omega _{x}$ and $\omega _{\perp }$ are respectively the
axial and radial frequencies of the magnetic trap. The last term
represents the potential due to the presence of the optical
lattice, with $\lambda $ denoting the wavelength of the
retroreflected laser beam, and $s$ being a factor characterizing
the intensity of the optical potential which increases when
increasing the intensity of the laser beam. In addition,
$E_{R}=2\pi ^{2}\hbar ^{2}/(m\lambda ^{2})$ is the recoil energy
of an atom absorbing one photon. For the optical lattice created
by the retroreflected laser beam, $d=\lambda /2$ is the distance
between two neighboring wells induced by the optical lattice.

The optical potential can be approximated as $V_{eff}=\sum_{k}\frac{1}{2}m%
\widetilde{\omega }_{x}^{2}(x-kd)^{2}$ with $\widetilde{\omega }_{x}=2\sqrt{s%
}E_{R}/\hbar $ \cite{1Dlatt}. As in Ref.\cite{1Dlatt} we assume that for the
parameters appearing in the potential (\ref{potential}) the condition $%
\omega _{x}<<\omega _{\perp }<<k_{B}T<\widetilde{\omega }_{x}$ is satisfied.
Thus the magnetic trap is a slowly-varying 3D potential which confines the
atoms harmonically to overall cigar-shaped distribution. In the 1D optical
lattice the atoms are confined to an array of 2D planes. As a result the
whole condensate consists of an array of quasi-2D subcondensates like disks
in shelf. Such array of quasi-2D subcondensates has been realized in the
experiment\cite{1Dlatt}.

The investigation of the phase diffusion of the above-mentioned BEC can be
divided into two steps. The first step is to consider the phase diffusion of
a single disk-shaped BEC. For this aim we develop a general approach to
calculate the probability distribution of the phase for a condensate
confined in a quasi-2D harmonic trap. For the temperature far below the
critical temperature $T_{c}$, the condensate can be described by the order
parameter with a phase factor $\phi \left( t\right) $:

\begin{equation}
\Phi \left( {\bf r},t\right) =\Phi _{0}\left( {\bf r}\right) e^{-i\phi
\left( t\right) }.  \label{phase1}
\end{equation}
The differential of the phase takes the form

\begin{equation}
{d\phi =\mu \left( N_{0},T\right) dt/\hbar ,}  \label{phase2}
\end{equation}
where ${\mu \left( N_{0},T\right) }$ is the chemical potential of the
system. We see that the particle-number fluctuations of the condensate yield
fluctuations in the chemical potential, and hence lead to the phase
diffusion of the condensate. Assuming that the mean ground state occupation
number is $\left\langle N_{0}\right\rangle $, in the case of the
particle-number fluctuations $\delta N_{0}<<N_{0}$, the average phase of the
condensate is then given by

\begin{equation}
\phi \left( \left\langle N_{0}\right\rangle ,t\right) =\mu \left(
\left\langle N_{0}\right\rangle ,T\right) t/\hbar .  \label{averagephase}
\end{equation}
The phase diffusion of the condensate can be described by considering the
phase difference $\Delta \phi \left( t\right) =\phi \left( t\right) -\phi
\left( \left\langle N_{0}\right\rangle ,t\right) $. From Eqs. (\ref{phase2})
and (\ref{averagephase}), it is straightforward to obtain a differential
equation on \ $\Delta \phi \left( t\right) $:

\begin{equation}
\frac{d\Delta \phi \left( t\right) }{dt}=\frac{\partial \mu \left(
\left\langle N_{0}\right\rangle ,T\right) }{\partial \left\langle
N_{0}\right\rangle }\Delta N_{0}\left( t\right) /\hbar .  \label{diff}
\end{equation}
In the above expression, $\Delta N_{0}\left( t\right) $ represents the
fluctuations of the ground state occupation number around $\left\langle
N_{0}\right\rangle $ at time $t$. Different from the situation considered by
Jin et al\cite{JIN} where a small time-dependent perturbation is used
artificially to create selected collective excitations in a BEC, the
collective excitations considered here are created spontaneously from the
condensate due to quantum fluctuations. Thus $\Delta N_{0}\left( t\right) $
changes with the development of time due to the creations and annihilations
of various collective modes.

For the temperature far below the critical temperature, the collective
excitations spontaneously created from the condensate play a dominant role
in the fluctuations of the condensate \cite{XIONG}. We therefore pay
attention here to the phase diffusion contributed by the collective
excitations of the system. For a pure 2D Bose-condensed gas, the collective
mode is labelled by two parameters $\left\{ n,l\right\} $. Here, the
parameter $l$ labels the angular momentum of the excitation. In this case,
one has $\Delta N_{0}\left( t\right) =\Sigma _{nl}\Delta N_{nl}\left(
t\right) $. Assuming that $\Delta \phi _{nl}\left( t\right) $ represents the
phase diffusion due to the collective mode $nl$, we have

\begin{equation}
\frac{d\Delta \phi _{nl}\left( t\right) }{dt}=\frac{\partial \mu \left(
\left\langle N_{0}\right\rangle ,T\right) }{\partial \left\langle
N_{0}\right\rangle }\Delta N_{nl}\left( t\right) /\hbar .  \label{diffnl}
\end{equation}
Thus the overall phase fluctuations of the condensate read $\Delta \phi
\left( t\right) =\Sigma _{nl}\Delta \phi _{nl}\left( t\right) $.

In the above equation, $\left| \Delta N_{nl}\right| $ can be regarded as the
mean number of atoms corresponding to the collective mode $nl$. Based on the
Bogoliubov theory \cite{BOG}, $\left| \Delta N_{nl}\right| $ is given by

\begin{equation}
\left| \Delta N_{nl}\right| =\left( \int u_{nl}^{2}({\bf r})dV+\int
v_{nl}^{2}({\bf r})dV\right) f_{nl}+\int v_{nl}^{2}({\bf r})dV,
\label{bogoliubov}
\end{equation}
where $f_{nl}=1/\left( \exp \left( \beta \varepsilon _{nl}\right) -1\right) $
is the number of the collective excitations presenting at thermal
equilibrium, $\beta =1/(k_{B}T)$ ($T$ is temperature), and $\varepsilon
_{nl}=\hbar \omega _{nl}$ is the energy spectrum of the collective mode $nl$%
. In the above equation, the quantity $\int v_{nl}^{2}({\bf r})dV$ describes
the effect of quantum depletion, which does not vanish even at $T=0$ . In
Eq. (\ref{bogoliubov}), $u_{nl}({\bf r})$ and $v_{nl}({\bf r})$ are
determined by the following coupled equations:

$${
\left( -\frac{\hbar ^{2}}{2m}\bigtriangledown ^{2}+V_{\perp}\left( {\bf r}%
\right) -\mu +2gn\left( {\bf r}\right) \right) u_{nl}({\bf r}) }$$

\[
{+gn_{0}\left( {\bf r}\right) v_{nl}({\bf r})=\varepsilon
_{nl}u_{nl}({\bf r}),}
\]

$${
\left( -\frac{\hbar ^{2}}{2m}\bigtriangledown ^{2}+V_{\perp }\left( {\bf r}%
\right) -\mu +2gn\left( {\bf r}\right) \right) v_{nl}({\bf r})
}$$

\begin{equation}
+gn_{0}\left( {\bf r}\right) u_{nl}({\bf r})=-\varepsilon
_{nl}v_{nl}({\bf r}), \label{uvenergy}
\end{equation}
where $n\left( {\bf r}\right) $ and $n_{0}\left( {\bf r}\right) $ are the
density distributions of the Bose gas and condensate, respectively. $%
V_{\perp }\left( {\bf r}\right) =\frac{1}{2}m\omega _{\perp
}^{2}(y^{2}+z^{2})$ is the 2D harmonic potential confining the Bose gas in $%
y $ and $z$ directions (the second term on the right hand side of (\ref
{potential})\thinspace ), while $g$ is the coupling constant describing the
interatomic interaction in the condensate.

By a straightforward calculation the excitation frequency of the collective
mode $nl$ is given by
\begin{equation}  \label{excspe}
\omega _{nl}=\omega _{\perp }\left( 2n^{2}+2n\left| l\right| +2n+\left|
l\right| \right) ^{1/2},
\end{equation}
while $\left| \Delta N_{nl}\right| $ reads

\begin{equation}
\left| \Delta N_{nl}\right| =\frac{\left( N_{0}mg\right) ^{1/2}\beta _{nl}}{2%
\sqrt{\pi }\hbar \sqrt{2n^{2}+2n\left| l\right| +2n+\left| l\right| }},
\label{nnnl}
\end{equation}
where
\begin{eqnarray}
& & \beta _{nl}=\int_{0}^{1}\left( 1-x^{2}\right) x\left( H_{nl}\right)
^{2}dx/\int_{0}^{1}x\left( H_{nl}\right) ^{2}dx,  \label{betanl} \\
& & H_{nl}\left( x\right)=x^{|l|}\sum_{j=0}^{n}b_{j}x^{2j}  \label{hnl}
\end{eqnarray}
with $b_{0}=1$ and the coefficients $b_{j}$ satisfying the recurrence
relation $b_{j+1}/b_{j}= \left( 4j^{2}+4j+4j|l|-4n^{2}-4n-4n|l|\right)
/\left( 4j^{2}+4j|l|+8j+4|l|+4\right) $.

As mentioned above, the collective excitations considered here are created
spontaneously due to the quantum fluctuations of the condensate. They can
disappear as time develops and hence have a finite longevity. The longevity
can be estimated based on the time-energy uncertainty relation. For the
collective mode $nl$, its longevity is $\tau _{nl}=1/\omega _{\perp }\left(
2n^{2}+2n\left| l\right| +2n+\left| l\right| \right) ^{1/2}$. Thus at time $%
t>>\tau _{nl}$, there is a series of $i=t/\tau _{nl}$ successive
creations and annihilations of the collective mode $nl$. Assuming
that at the time $t$ the number of the collective modes created
(annihilated) in the condensate is $i_{cre}$ ($i_{ann}$). Thus one
has $i=i_{cre}+i_{ann}$. In this situation, the desired
probability is given by the binomial expression \cite{PATHRIA}

\begin{equation}
P_{i}\left( j\right) =\left( \frac{1}{2}\right) ^{i}\frac{i!}{\left\{ \frac{1%
}{2}(i+j)\right\} !\left\{ \frac{1}{2}(i-j)\right\} !},  \label{binomial}
\end{equation}
where $j=i_{cre}-i_{ann}$. For $t>>\tau _{nl}$, using the Stirling's
formula, the asymptotic form of the probability distribution $P_{nl}\left(
\Delta \phi _{nl},t\right) $ of the phase diffusion due to the collective
mode $nl$ takes the form

\begin{equation}
P_{nl}\left( \Delta \phi _{nl},t\right) =\frac{1}{\Gamma _{nl}\sqrt{2\pi
t/\tau _{nl}}}\exp \left( -\frac{\tau _{nl}\left( \Delta \phi _{nl}\right)
^{2}}{2\Gamma _{nl}^{2}t}\right) ,  \label{probnl}
\end{equation}

\vspace{5mm}

\noindent where the dimensionless parameter $\Gamma _{nl}$ is
given by

\begin{equation}
\Gamma _{nl}=\frac{\partial \mu \left( \left\langle N_{0}\right\rangle
,T\right) }{\partial \left\langle N_{0}\right\rangle }\frac{\left| \Delta
N_{nl}\right| \tau _{nl}}{\hbar }.  \label{ttnl}
\end{equation}
From the Gaussian distribution given by Eq. (\ref{probnl}), the phase
fluctuations due to the collective mode $nl$ have the form:

\begin{equation}
\left( \delta ^{2}\phi \right) _{nl}=\Gamma _{nl}^{2}t/\tau _{nl}.
\label{flucnnll}
\end{equation}

Assuming that the collective modes with different $nl$ are created and
annihilated independently, one obtains that the probability distribution of
the overall phase difference $\Delta \phi \left( t\right) $ is still a
Gaussian distribution function. The probability distribution of $\Delta \phi
$ is then given by

\begin{equation}
P\left( \Delta \phi ,t\right) =\sqrt{\frac{B}{\pi }}\exp \left( -B\left(
\Delta \phi \right) ^{2}\right) .  \label{probability1}
\end{equation}
Based on the general probability theory, the parameter $B$ can be obtained
through the relation $\delta ^{2}\phi =1/2B=\Sigma _{nl}\left( \delta
^{2}\phi \right) _{nl}$. From the formulas (\ref{nnnl}), (\ref{ttnl}), (\ref
{flucnnll}) and the chemical potential $\mu \left( N_{0},T\right) =\left(
N_{0}mg/\pi \right) ^{1/2}\omega _{\perp }$ for 2D harmonic trap, we obtain
the overall phase fluctuations $\delta ^{2}\phi $ as

\begin{equation}
\delta ^{2}\phi =\frac{m^{2}g^{2}\omega _{\perp }t}{4\pi ^{2}\hbar ^{4}}%
\Upsilon ,  \label{totalphase1}
\end{equation}
where the coefficient $\Upsilon $ takes the form

\begin{equation}
\Upsilon =\sum_{nl}\frac{\beta _{nl}^{2}}{\left( 2n^{2}+2n\left| l\right|
+2n+\left| l\right| \right) ^{3/2}}\left[ f_{nl}+\frac{1}{2}\right] ^{2},
\label{coefficient}
\end{equation}
where $\beta_{nl}$ has been given in Eq. (\ref{betanl}). From Eq. (\ref
{totalphase1}) we see that when the longevity of the collective excitations
is taken into account, the phase fluctuations are proportional to the time $%
t $, rather than $t^{2}$ which is obtained when the collective excitations
are regarded as quite stable.

As mentioned above, if $\omega _{x}<<\omega _{\perp }<<k_{B}T<\widetilde{%
\omega }_{x}$, the condensate confined in each well has a quasi-2D nature.
Note that for the array of the subcondensates induced by the optical lattice
with enough depth, the overlap between the subcondensates occupying
different wells can be omitted. This means that the behavior of the
condensate in each well has a local property. For each well the coupling
constant $g$ is given by $g\approx 2\sqrt{2\pi }\hbar ^{2}a_{s}/(m\widetilde{%
l_{x}})$ \cite{PETROV}, which is fixed by a $s$-wave scattering length $%
a_{s} $ and the oscillator length $\widetilde{l_{x}}=\left( \hbar /m%
\widetilde{\omega }_{x}\right) ^{1/2}$ in the $x$-direction. Thus from Eq. (%
\ref{totalphase1}) the subcondensate in each well displays the phase
fluctuations

\begin{equation}
\delta ^{2}\phi =\frac{2\Upsilon \omega _{\perp }t}{\pi }\left( \frac{a_{s}}{%
\widetilde{l_{x}}}\right) ^{2}.  \label{totalphase2}
\end{equation}
From the above result, one has $\delta ^{2}\phi \sim \omega _{\perp }%
\widetilde{\omega }_{x}$. We see that the confinement induced by the optical
lattices has the effect of increasing the phase fluctuations. This is a
natural character by considering the fact that the confinement has the
effect of increasing the particle-number fluctuations \cite{XIONGFLU}.

\section{Effect of phase diffusion on the interference pattern of expanding
array of subcondensates}

We now go to the second step, i. e. to discuss the phase diffusion of an
array of subcondensates confined in the combined potential of the
harmonically trapping potential and 1D optical lattice. To illustrate
clearly the role of the phase diffusion in the interference pattern of the
expanding subcondensates, we consider here an experiment scheme which can be
realized in future experiment. Firstly, a cigar-shaped condensate is formed
in a magnetic trap with the harmonic angular frequencies $\omega _{x}=$ $%
2\pi \times 9$ ${\rm Hz}$ and $\omega _{\perp }=2\pi \times 92$ ${\rm Hz}$
at a temperature far below the critical temperature. The atoms are
transferred into the lattice potential by increasing the power of the laser
beam so that the harmonic angular frequency of the well is approximated to
be $2\pi \times 14$ ${\rm kHz}$ in $x-$direction. For $\widetilde{\omega }%
_{x}=2\pi \times 14$ ${\rm kHz}$, based on the well-known Bose-Hubbard
model, the tunnelling time between neighboring subcondensates is of the
order of $1$ {\rm ms}. In this situation, the array of subcondensates should
be regarded as fully coherent and there is no independent random phase for
the subcondensates in different wells. Although there is still a phase
diffusion for the whole condensate, one can not observe this through the
interference pattern. To observe the effect of the phase diffusion, one can
rapidly increase the lattice potential depth to a value of $\widetilde{%
\omega }_{x}=$ $2\pi \times 52$ ${\rm kHz}$ within a time (for example $50$ $%
{\rm \mu s}$) much smaller than the tunnelling time. In this
situation, the phase of every subcondensate is the same after the
lattice
potential ramps to its final strength. For the lattice potential with $%
\widetilde{\omega }_{x}=$ $2\pi \times 52$ ${\rm kHz}$, the tunnelling time
between neighboring wells is of the order of $1.5$ ${\rm s}$. Thus, for the
holding time $t_{h}$ of the optical lattice being much smaller than $1.5$ $%
{\rm s}$, the phase diffusion of the subcondensates in different wells can
be regarded as independent from each other. In this situation, there is a
random phase for each subcondensate with the development of the holding time
and this will lead to an obvious effect on the interference pattern after
the combined potential is switched off.

For the experimental parameters considered here, we have $\delta ^{2}\phi
=1.60\times t_{h}$. Based on the result given by Eq. (\ref{totalphase2}),
the probability distribution of the phase for the subcondensate in each well
can be obtained through Eq. (\ref{probability1}). For the interference
pattern to wash out completely, the time scale of the holding time is
obtained by using $\delta ^{2}\phi =\pi ^{2}$. In the experimental
parameters used here, for the subcondensate in each well $\delta ^{2}\phi $
is still much smaller than $\pi ^{2}$ even when the holding time $t_{h}$ of
the optical lattice is $1{\rm s}$. Due to the fact that there are many
subcondensates induced by the optical lattice, however, we will show that
there is a significant effect of the phase diffusion to the interference
pattern after the optical lattice is switched off.

To observe a notable effect of the phase diffusion, we consider the case
when only the optical lattice is switched off after the holding time $%
t_{h}=0.2$ ${\rm s}$. Assume $t_{0}$ is the time after the optical lattice
is switched off. For the subcondensate confined in the $k$th well of the
optical lattice, applying the propagator method used in Ref.\cite{XIONG2} it
is easy to get the normalized wave function $\varphi _{k}\left(
x,t_{0}\right) $ after only the optical lattice is switched off:

\[
{\varphi }_{k}{\left( x,t_{0}\right) =A_{n}\sqrt{\frac{1}{\sin
\omega _{x}t_{0}\left( {\rm ctg}\omega _{x}t_{0}+i\gamma \right)
}}\exp \lbrack iR_{k}\left( \delta \phi ,t_{h}\right) \rbrack  }
\]

$${
\times \exp \left[ -\frac{\left( kd\cos \omega _{x}t_{0}-x\right)
^{2}}{2\sigma ^{2}\sin ^{2}\omega _{x}t_{0}\left( {\rm
ctg}^{2}\omega _{x}t_{0}+\gamma ^{2}\right) }\right]
 }$$

$${
\times \exp \left[ -\frac{i\left( kd\cos \omega _{x}t_{0}-x\right) ^{2}{\rm %
ctg}\omega _{x}t_{0}}{2\gamma \sigma ^{2}\sin ^{2}\omega
_{x}t_{0}\left( {\rm ctg}^{2}\omega _{x}t_{0}+\gamma ^{2}\right)
}\right] }$$

\begin{equation}
\times\exp \left[ \frac{%
i\left( x^{2}\cos \omega _{x}t_{0}+k^{2}d^{2}\cos \omega
_{x}t_{0}-2xkd\right) }{2\gamma \sigma ^{2}\sin \omega _{x}t_{0}}\right] {,}
\label{evolution2}
\end{equation}
where $A_{n}=1/\pi ^{1/4}\sigma ^{1/2}$ is a normalization constant, the
dimensionless parameter $\gamma =\hbar /m\omega _{x}\sigma ^{2}$ with $%
\sigma $ the width of the subcondensate in each well \cite{XIONG2}. The
factor $R_{k}\left( \delta \phi ,t_{h}\right) $ represents a random phase
due to the phase diffusion of the subcondensate in the $k$th well. In
numerical calculations, $R_{k}\left( \delta \phi ,t_{h}\right) $ is
generated according to the probability distribution function $P\left( \Delta
\phi ,t_{h}\right) $ (see (\ref{probability1})\thinspace ). Then we obtain
the density distribution in $x-$direction as follows:

\begin{equation}
n\left( x,t_{0}\right) =\Xi \left| \sum_{k=-k_{M}}^{k_{M}}{\left( 1-\frac{%
k^{2}}{k_{M}^{2}}\right) \varphi _{k}}{\left( x,t_{0}\right) }\right| ^{2},
\end{equation}
where $\Xi =15N_{all}k_{M}^{3}/\left( 16k_{M}^{4}-1\right) $. $N_{all}$ is
the total number of atoms in the array of condensates and $2k_{M}+1$ is the
total number of the subcondensates induced by the optical lattice. In the
present work, by using the experimental result in \cite{1Dlatt}, $k_{M}$ is
chosen to be $100$.

It is obvious that the density distribution $n\left( x,t_{0}\right) $ at $x=0
$ reaches the maximum value at time $t_{l}=\left( 2l-1\right) \pi /2\omega
_{x}$, where $l$ is a positive integer. Displayed in Fig. 1 is $n\left(
x,t_{0}\right) $ (with $N_{all}\Xi A_{n}^{2}$ as a unit) at $t_{l}$ with the
holding time $t_{h}=0.2$ ${\rm s}$. We see that the phase diffusion of the
array of the subcondensates makes the central density of the interference
pattern decrease significantly. Displayed in Fig. 2 is the density
distribution $n\left( x=0,t_{0}=t_{l}\right) $ versus the holding time $t_{h}
$. A exponential damping of the central density is clearly shown in the
figure.

\begin{figure}[tb]
\psfig{figure=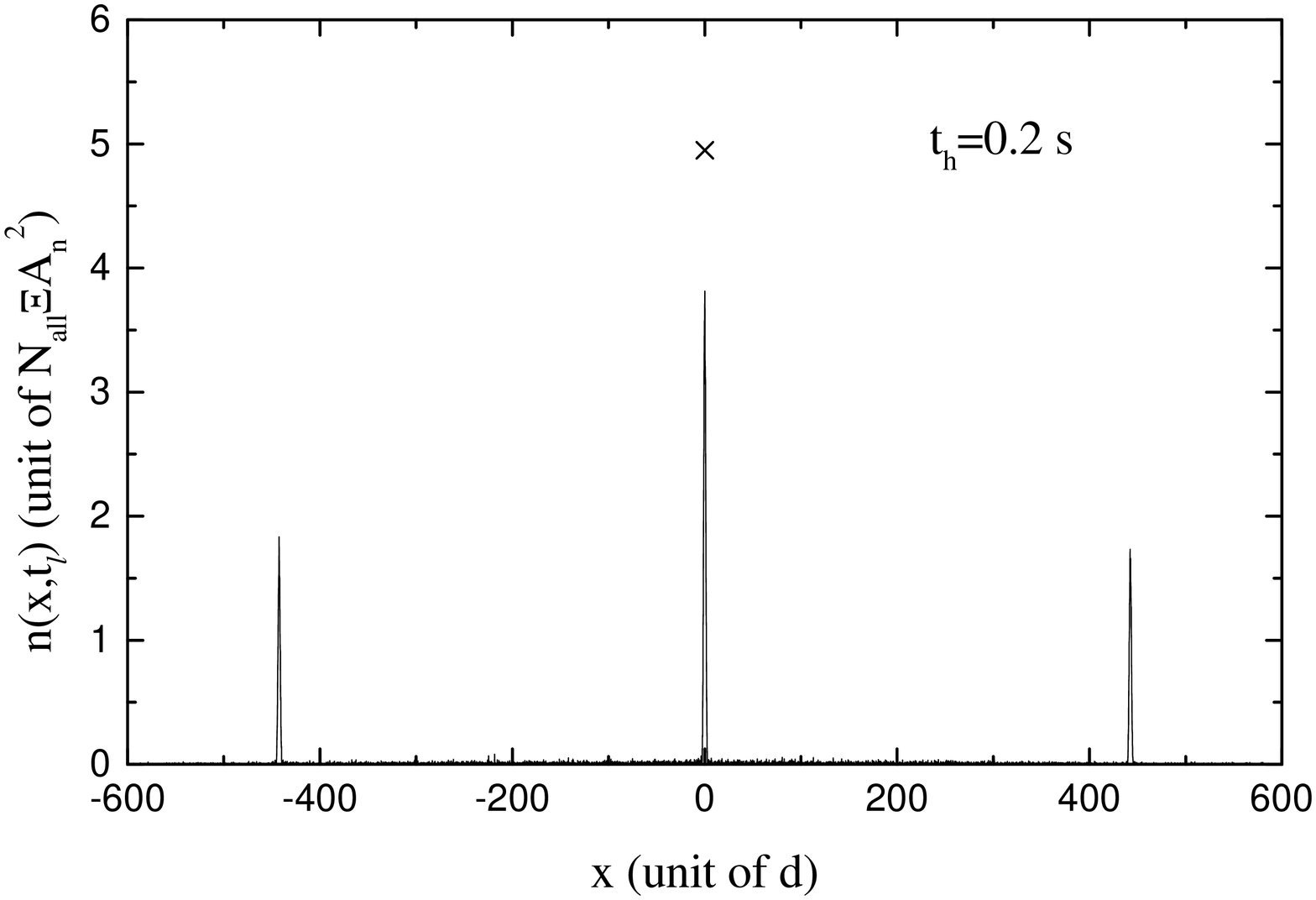,width=\columnwidth,angle=270}
\caption{Displayed is the density distribution of the interference pattern at $%
t_{0}=t_{l}$ for the case of the holding time $t_{h}$ of the
optical lattices being $0.2$ ${\rm s}$. The density distribution
$n\left(
x,t_{l}\right) $ is in unit of $N_{all}\Xi A_{n}^{2}$, while the coordinate $%
x$ is in unit of $d$. In this figure, the position of the cross
$\times $ indicates the density at $x=0$ in the case of $t_{h}=0$
${\rm s}$. The figure shows clearly that the phase diffusion has
the effect of decreasing significantly the maximum density of the
central peak.}
\end{figure}

\begin{figure}[tb]
\psfig{figure=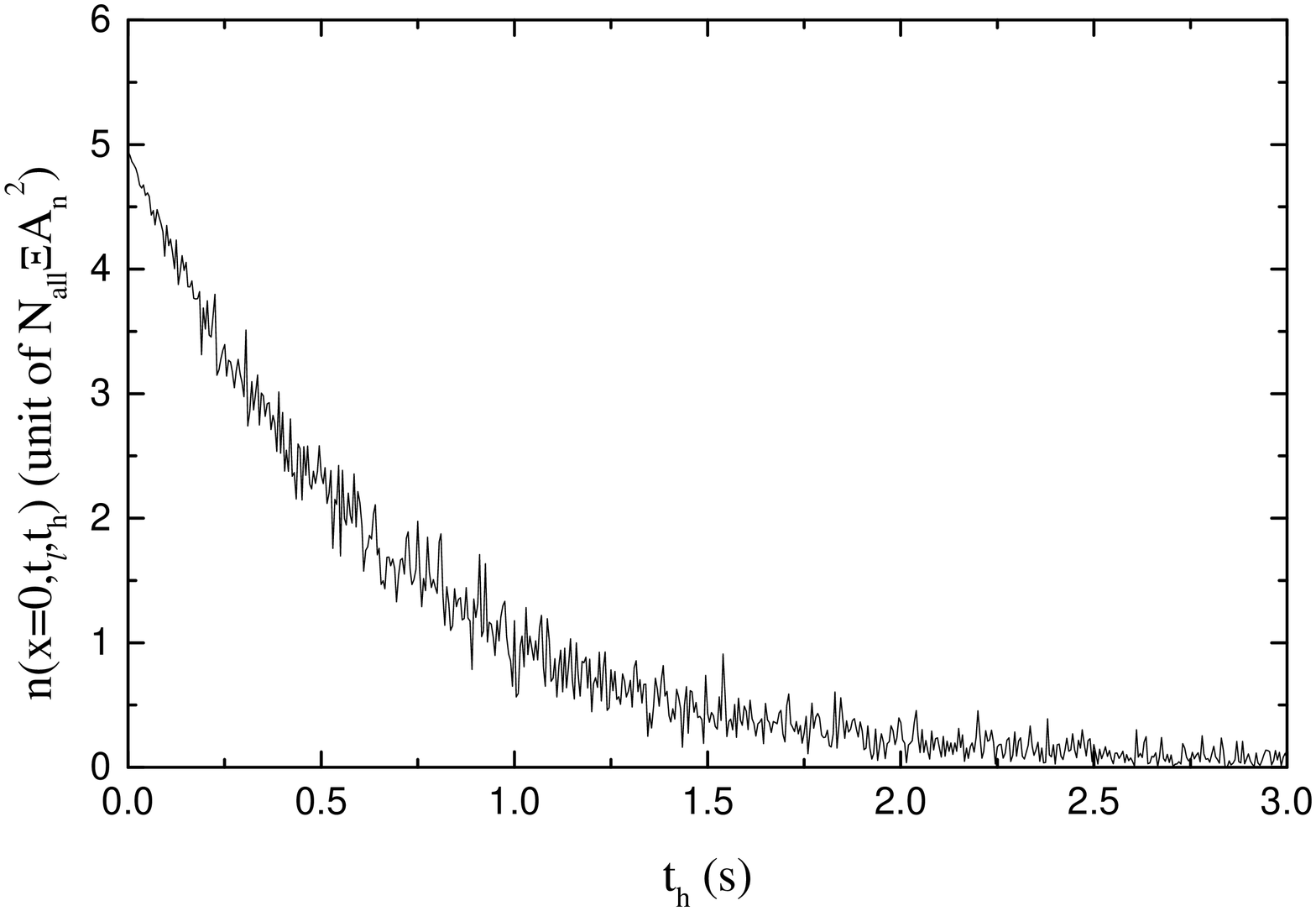,width=\columnwidth,angle=270}
\caption{Shown is $n\left( x=0,t_{l},t_{h}\right) $ versus the holding time $%
t_{h}$. The exponential damping of the central peak is clearly
shown in the figure.}
\end{figure}

It is straightforward to calculate the interference pattern of the expanding
array of BECs when the magnetic trap and optical lattice are both
switched off. Shown in Fig. 3 is the interference pattern for the holding
time $t_{h}=0.2$ ${\rm s}$ after the combined potential is switched off for $%
30$ ${\rm ms}$. In comparison with the case of $t_{h}=0$ ${\rm s}$ (dashed
line in Fig. 4), the strong noise in Fig. 3 shows that the phase diffusion
has an important effect on the interference pattern of the array of
subcondensates. The solid line in Fig. 4 displays the average density
distribution of twenty ensembles for the holding time $t_{h}=0.2$ ${\rm s}$.
We see that the maximum density of the central and side peaks decreases
significantly due to the presence of the phase diffusion. Generally
speaking, there are four parameters $\omega _{\perp }$, $\widetilde{\omega }%
_{x}$, $k_{M}$, $t_{h}$ which are related closely to the effect of the phase
diffusion. The present work shows that increasing these parameters will
contribute to the observation of the phase diffusion for BEC in optical
lattices.

\begin{figure}[tb]
\psfig{figure=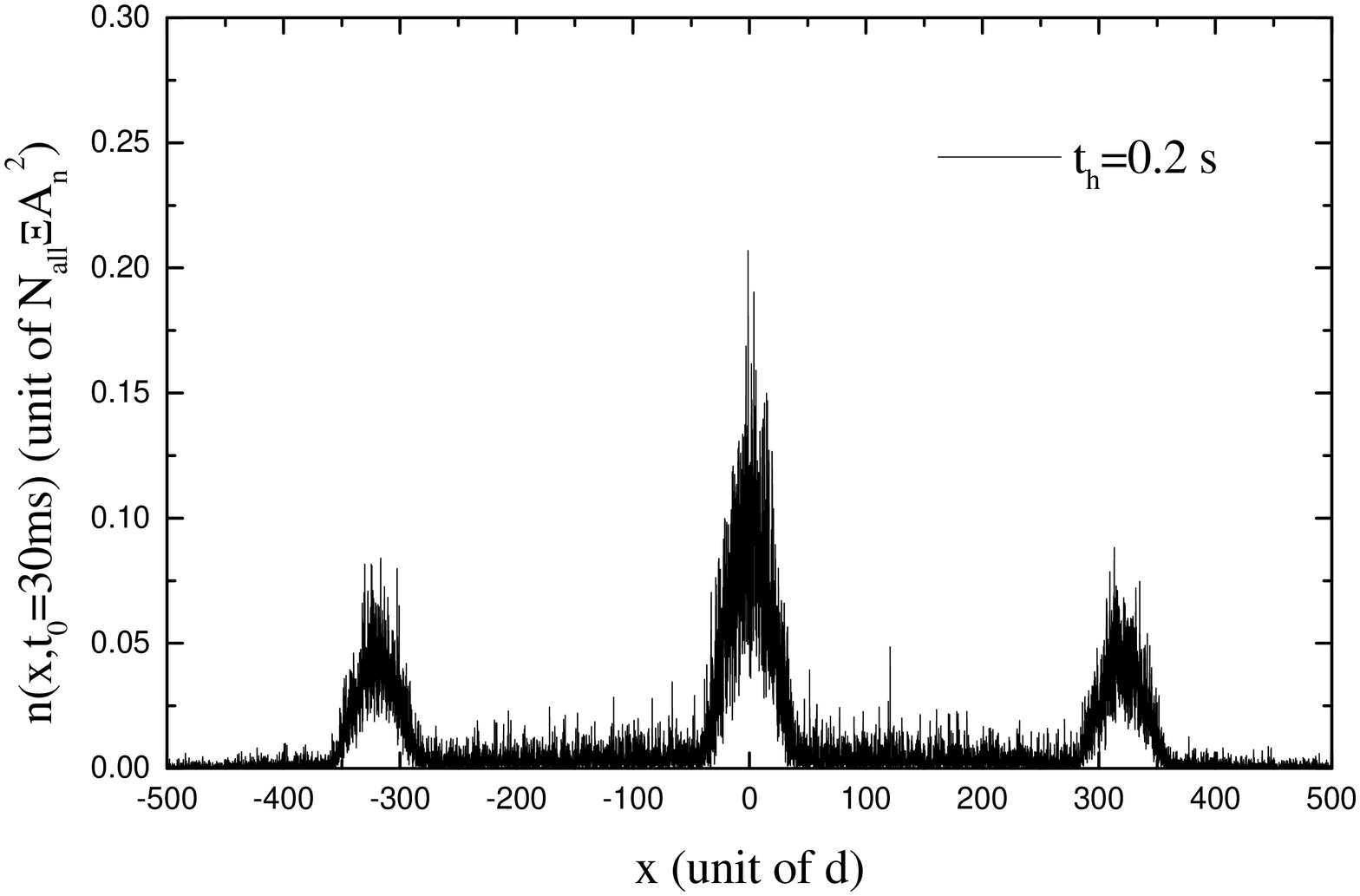,width=\columnwidth,angle=270}
\caption{After the holding time $0.2$ ${\rm s}$ of the optical
lattices, when both the magnetic trap and optical lattices are
switched off, the density
distribution of the expanding condensates is shown for $t_{0}=30$ ${\rm ms}$%
. In the figure, $n\left( x,t_{l}\right) $ is in unit of
$N_{all}\Xi A_{n}^{2}$, while $x$ is in unit of $d$. Compared with
the density distribution of the zero holding time (dashed line in
Fig. 4), the effect of the phase diffusion on the interference
pattern is very obvious.}
\end{figure}

\begin{figure}[tb]
\psfig{figure=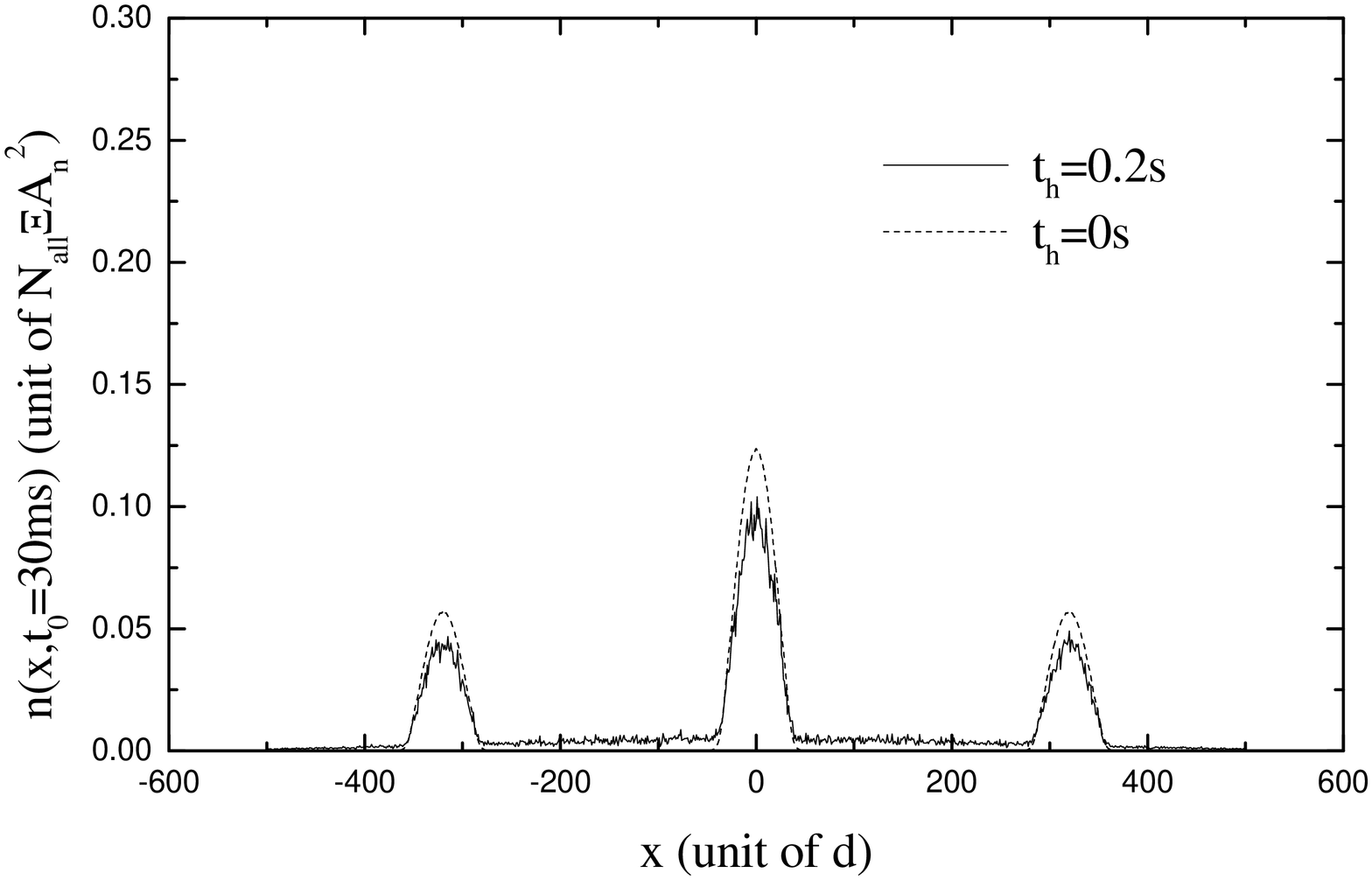,width=\columnwidth,angle=270} \caption{The
solid line displays the average density distribution of twenty
ensembles when both the magnetic trap and optical lattices are
switched off
for $t_{0}=30$ ${\rm ms}$ in the case of the holding time $t_{h}=0.2$ ${\rm s%
}$. The dashed line shows the interference pattern in the case of
the holding time $t_{h}=0$.}
\end{figure}

\section{Discussion and summary}

Based on the analysis of the collective excitations, we have investigated
the phase diffusion at low temperature in an array of subcondensates formed
in a combined potential consisting of harmonically magnetic trap and 1D
optical lattice. We have shown that the effect of phase diffusion on the
interference pattern is evident and thus can be measured from the expanding
subcondensates after the optical lattice is switched off. We point out that
the maximum density of the interference pattern decreases significantly due
to the mixing of the phase diffusion appearing in different subcondensates
formed by the optical lattice. This suggests a way to detect experimentally
the notable phase diffusion of BEC. For the case of BEC\ in 3D optical
lattice, however, it seems that the theory developed here can not be used
due to the fact that there are only an average atom number of up to $2.5$
atoms per lattice site \cite{GREI}.

Note that a recent experiment has been conducted in Ref.\cite{MORSCH} where
the interference pattern is measured when only the optical lattice is
switched off. Clearly this type of experiment can be used to test the
theoretical predictions provided in this paper. We stress that for a single
subcondensate the phase fluctuations increase very slowly with the
development of the time. However, the phase diffusion plays an important
role in the interference pattern of the expanding subcondensates, as shown
in the present work.

%%%%%%%%%%%%%%%%%%%%%%%%%%%

\section*{Acknowledgments}

%%%%%%%%%%%%%%%%%%
We acknowledge valuable discussions with L. You. This work was supported by
Natural Science Foundation of China under grant Nos. 10205011 and 10274021.

\end{document}